\RequirePackage{fix-cm}
\documentclass[a4paper, 12pt]{article} 
\usepackage{graphicx, color}
\usepackage[utf8]{inputenc}
\usepackage[T1]{fontenc}
\usepackage{amsmath, makecell, booktabs, caption}

\usepackage{cite}
\usepackage{colortbl}
\usepackage{hyperref}

\definecolor{linkcolor}{rgb}{0.9,0,0}
\definecolor{citecolor}{rgb}{0,0.6,0}
\definecolor{urlcolor}{rgb}{0,0,1}
\hypersetup{
	colorlinks, linkcolor={linkcolor},
	citecolor={citecolor}, urlcolor={urlcolor}
}

\newcommand{\ket}[1]{\left| #1\right\rangle}
\newcommand{\braket}[2]{\left\langle #1\right|\!\!\left.#2\right\rangle}

\begin{document}

\title{Quantum Cognitive Triad: \\ Semantic geometry of context representation}


\author{Ilya A. Surov}



\maketitle

\begin{abstract}
	
The paper describes an algorithm for semantic representation of behavioral contexts relative to a dichotomic decision alternative.
The contexts are represented as quantum qubit states in two-dimensional Hilbert space visualized as points on the Bloch sphere. The azimuthal coordinate of this sphere functions as a one-dimensional semantic space in which the contexts are accommodated according to their subjective relevance to the considered uncertainty.
The contexts are processed in triples defined by knowledge of a subject about a binary situational factor. The obtained triads of context representations function as stable cognitive structure at the same time allowing a subject to model probabilistically-variative behavior.
The developed algorithm illustrates an approach for quantitative subjectively-semantic modeling of behavior based on conceptual and mathematical apparatus of quantum theory.

\end{abstract}

\section{Introduction}
\label{intro}

Science begins when one starts to quantify. This truth underlying metrological work of D.I. Mendeleev \cite{Mendeleev1950} is relevant for current understandings of behavioral and cognitive motivations \cite{Freud1923,Adler1923}, traits of perception and thinking \cite{James1890,Young2016}, classification of personalities \cite{Jung1921,Bukalov2005} and systematization of unconscious cognition \cite{Jung2014,Westen1999,Augusto2010a} which do not imply any quantitative measure. 
This is typical for a <<preparadigmatic>> state of psychological science distinguished by separated research and practice, divergent conceptual orientations, and little consent on foundational terms \cite{Melchert2016,Goldfried2019,Zagaria2020}.

Metrological deficiency of psychology scales to the collective level. Even when certain agreement on terms and procedures is reached, repeatability of experiments, as compared to physics, is strikingly low \cite{OSC2015,Camerer2018}. Development of practical psychology, reliable social and economical engineering in this scientific paradigm is therefore recognized as problematic \cite{Ball2006,Bouchaud2008,Krugman2009,Lilienfeld2012,Ferguson2015,Smedslund2016}. This situation of antagonism between <<humanities and sciences>> is recognized as a call for development of integrative approach linking these domains of knowledge \cite{MEXT2015}.
Akin to diverse families of archaic measures of lengths, volumes and weights revised by Mendeleev, contemporary concepts of psychology and sociology are to be reconciled with measuring procedures and quantitative terms.


\subsection{Classical behaviorism}
\label{clb}

Quantification of humanitarian sciences was attempted by Pavlov, Sechenov, Watson and Fechner who renounced introspective study of their own <<consciousness>> in favor of modeling behavior of their neighbors. The plan was to formalize psychology with the same scientific method which turned ancient philosophy, medieval astrology and alchemy to their contemporary counterparts. 
Seemingly reasonable, this strategy did not produce for psychology a reliable theoretical structure comparable to physics or chemistry \cite{Young2016}. The reason is that classical behaviorist models view a human being as a mechanical automaton programmed to execute a set of stimulus-response scripts; no room for creativity and free will in these models leaves higher psychological functions on the ground of verbal descriptions criticized above.

Stimulus-response structure of classical behaviorism is generalized by supplementing it with a decision-making agency which results in a stimulus-organism-response scheme \cite[ch.28]{Young2016}. Still, within classical methodology this approach implies specification of an internal machinery of the <<organism>>; predictably, this leads to mechanistic models of behavior again throwing a baby out with the bathwater.

\paragraph{Methodology} of classical behaviorism aims to decompose an object into <<elementary>> parts and then, starting from a heap of those pieces, strives to assemble the whole thing back. In practice, though, the latter process depends on one's ability to recover relations between the components which may be lost or destroyed in the former one \cite{Maturana1995a}.
While effective for analysis of complex inert systems like fine electronics and megapolis infrastructure, when applied to living systems this approach faces difficulties at modeling of even a single living cell \cite{Breuer2019}; for example, the behavior of a microscopic worm whose three hundred neurons and several thousand couplings were perfectly known three decades ago is yet to be understood \cite{Larson2018,Cook2019}. Acknowledging contemporary progress of neuronal imaging \cite{Glasser2016}, one is bound to recognize that the understanding of human behavior based on a detailed neural reconstruction is far from practical results \cite{Stix2013}.

\subsection{Quantum approach}
\label{adventQuantum}

\paragraph{Historical context}
The reason for the difficulties encountered by classical behaviorism was hypothesized at the end of 1970' when nuclear weapons together with the transistor and laser technologies entered public cognition \cite{Gurevich1977,Orlov1981,Orlov1982}. These new capabilities displayed the power of quantum physics – a novel branch of science with conceptual structure radically different from anything known in natural science before \cite{Bohr1933,Wheeler1989}, at the same time expressed in the mathematical language producing quantitative models of fabulous precision. 
This newly certified theory of atomic-scale phenomena appeared as an alternative to the methodology of Newtonian mechanics which was not available to the founders of classical behaviorism. 

\paragraph{Progress}
Discontent about classical humanitarian science both on individual and collective levels \cite{Kjellman2006,Ball2006,Bouchaud2008,Ferguson2015} motivated a search for new paradigms of behavioral modeling. In this respect quantum theory was recognized as a conceptual and mathematical system, applicability of which extends beyond physics \cite{Aerts1995,Atmanspacher2002}.
Used as a generalization of classical probability calculus to model cognition and behavior of humans, quantum theory showed efficiency in the areas problematic for other approaches including irrational preference, contextual decision making and non-expected game equilibria, modeling of natural language, collective cognitive and behavioral excitations. These and other topics are covered in the reviews \cite{Busemeyer2011,Agrawal2013,Khrennikov2015c,Bruza2015,Asano2015} and monographs \cite{Khrennikov2010,Busemeyer2012,Haven2013,Asano2015b}.

\paragraph{Quantum language}
Quantum cognitive science exploits correspondence between phenomena of human behavior and mathematically expressed concepts of quantum theory, most productive of which are state space, superposition, entanglement, observable, and measurement. In short, quantum models consider a human individual in a particular behavioral context as an instance of a physical system prepared in a particular quantum state which encodes available decision alternatives and propensities of their realization \cite{Agrawal2013}. The latter are quantified by complex-valued amplitudes which contrasts to the classical measure of uncertainty in terms of real-valued probabilities.

\paragraph{The phase problem}

The phase dimension of complex-valued amplitudes, crucial in many quantum models of cognition and behavior, is mostly used as an additional parameter meaning of which is usually undisclosed. 
This renders many quantum models of cognition and behavior to the role of post-factum fitting apparatus incapable of predictive modeling.

This problem arises from a general feature of quantum theory according to which phase parameters do not enter directly in real-valued observable probabilities and, therefore, have no straight measurement procedure \cite{Lynch1995}.
Nevertheless, in electron spin states, photon polarization, and other elementary systems meaning of phases was successfully guessed and validated in experiments which allowed to develop techniques for phase-sensitive reconstruction of the corresponding quantum states \cite{Rehacek2004}. 
The accomplishment of a similar task in quantum science of cognition and behavior of living systems would tremendously increase the practical value of this field.

\paragraph{This paper} suggests an understanding of the phase parameter in a basic theoretical structure of quantum cognitive science - the qubit. 
This understanding follows from a specific use of the quantum state as a subjective representation of behavioral context explained in Sect.~\ref{method} and realized in a theoretical structure called quantum cognitive triad developed in Sect.~\ref{model}.
The developed model results in an algorithm for context representation which is tested in Sect.~\ref{experiment} on the experimental data collected from several studies of a two-stage gambling task. 
The model performance is discussed in Sect.~\ref{semantics}. 
Sect.~\ref{outlook} outlines the implications of the result.

\section{Methodology: quantum behaviorism}
\label{method}

Instead of breaking things apart, quantum methodology aims to model only the behavior of the whole system, while its internal mechanisms may remain unknown. Compared to the classical criterion of scientific knowledge, this amounts to a significant decrease in ambition. Such a retreat was not accepted easily; physicists turned to this humble stance after decades of a fruitless fight for the classical understanding of what is going on in the quantum labs around the world \cite{Bohr1933,Wheeler1989,Wiseman2015}. 
This seemingly weak position, however, constitutes a crucial advantage of the quantum methodology in application to humanities.

\subsection{Behavior and the black box}
\label{bbb}

The problem addressed by quantum behaviorism is the probabilistic modeling of the multivariant processes in nature.
The process is a behavior of a subject system - an entity revealing itself to the outside exclusively through the observable decisions \cite{Chuang1997}.
Contrary to physics, human-centered sciences invite this <<black box>> approach naturally since anyone else's mind is usually accessed not directly but via observable signs \cite{Heyes2014,Salvatore2020}.

Quantum cognition faces this fact head-on. It does not seek to differentiate between thoughts, motives, emotions, moods, and tempers.
An enormous complexity of the human psyche is beyond the experimenter's control, but not neglected; included in the black box, it is free to affect the observable behavior and find reflection in quantum models. 
Focus on the observables and their separation from the interpretative constructs relieves quantum approach from the quantification problems encountered in psychology \cite{Stevens1946,Toomela2008,Tafreshi2016,Kostromina2019} (further discussion of this in Sect.~\ref{reflexivity}).

\subsection{Discrimination and alternatives}
\label{alternatives}

The quantum model formalizes behavior of the black box as a choice among a set of distinguished alternatives, e.g. turning left or right on a crossroad. This definition ignores countless features of a particular performance within either option which are not discriminated as distinct behavioral alternatives\footnote{Such coarse-graining of observables considered as a fundamental premise of quantum theory \cite{Kofler2007} conforms with a discrete representation of information in the human's brain \cite{Tee2020}.}. 

An elementary example of the discrete behavior is a deflection of an electron entering an inhomogeneous magnetic field; in this experiment of Stern and Gerlach the number of possible alternatives is exactly two, namely to deflect either along or opposite to the field's gradient \cite{Feynman1964}. 
The same kind of dichotomic alternative appears before a living organism when a situation forces it to choose between yes or no, up or down, doing or not doing.
As in physics, these alternatives are inherent neither to the black box nor to the experimental apparatus alone, but to their relation: possible choices are defined in a particular situation where one subject may distinguish not two but three behavioral options, whereas another individual may not recognize optionality at all.

\subsection{Objective and subjective uncertainty}
\label{uncertainty}

The very concept of behavior implies the possibility of a subject to choose between the alternative scenarios of action; behavioral modeling is therefore inherently probabilistic. Estimation of decision probabilities requires a set of identical experiments (i.e. indistinguishable in the sense of Sect.~\ref{alternatives}) performed on instances of identically prepared black boxes.
Each experiment in this series resolves an instance of behavioral uncertainty, which can be of two different types.


\paragraph{Example}
The first type of uncertainty is exemplified by the length of a book that is not yet finished; before the text is completed this (observable) length does not exist. This is an objective uncertainty that can only be resolved by an actual change of the system: the author has to write, choosing linguistic forms and possibly modifying ideas in the process.
Once the book is finished, the uncertainty is resolved: the number of the words and pages is recorded in nature and objectively exists. Still, even the author may not know these numbers if he or she did not count while writing.
This ambiguity conditioned by personal ignorance of a subject illustrates the second type of uncertainty.

\paragraph{Generally}, subjective uncertainty accounts for the ignorance of some person about an actual state of the system which predetermines the resolution of such uncertainty. This is done by a measurement that rewrites the information from an actual state of the system to the subject's brain via intermediate representations determined by the peripheral nervous system, possibly extended with a measurement apparatus.
Subjective uncertainty is modeled by the classical probability theory including the Bayes' rule accounting for modification of the subjective uncertainty due to update of the information available to the subject \cite{Kolmogorov1956}.

Quantum uncertainty, in contrast, accounts for objectively multivariant future of the considered system or process. Objective uncertainty may be resolved only by an actual transfer of the system to a state in which the considered alternative takes definite value. This process, ambiguously denoted by the same word <<measurement>>, is fundamentally different from its classical counterpart \cite{Bell1990}.
Objective uncertainty is modeled by quantum theory and is called quantum uncertainty, indeterminism or potentiality \cite{Bell1993,Gabora2005,Jaeger2012}.

\paragraph{Implications}

Objective multivariance of the living systems' behavior is a crucial ingredient missing in classical behaviorism. Even when the latter turns to a probabilistic view, the underlying Boolean algebra of events allows it to capture only those phenomena which follow a predetermined course like motion of the center of mass of a human body between jump and landing which does not differ in its behavior from a bag of sand.
Models of behavior based on the Boolean algebra or events and the Kolmogorovian probability calculus imply that their subjects have already taken every possible decision and thus only reproduce the fixed tables of the stimulus-response pairs; all possible quantities already have definite values so that no new information can be created.
As illustrated by Einstein, Podolsky and Rosen's famous setup \cite{Einstein1935}, this essentially static worldview agrees neither with quantum theory nor with the actual state of affairs \cite{Wiseman2015}. 


Allowing for objective potentiality, quantum theory constitutes a tool to model processes in which things happen and the future exists in the real sense\footnote{Cf. with the propensity interpretation of probability \cite{Popper1978a} and a tensed ontology of events described in \cite{Guarino2017}.}. This method of modeling does not seek to put the considered system in a condition where its behavior is predetermined; instead, quantum experiment explicitly focuses on objectively multivariant behavior of an intentionally uncontrolled subject - the black box.
Identification of the black box and definition of subjectively distinguished alternatives (Sect.~\ref{bbb} and \ref{alternatives}) constitutes the first step of quantum behavioral modeling\footnote{While consensus on the nature of creativity and free will is not reached, quantum models of behavior have space for phenomena of this kind \cite{Briegel2012,Stapp2017}.}.

\subsection{Contexts and Hilbert space}
\label{context}

In case of objective uncertainty, probabilities with which possible alternatives may actualize in the potential experiment depend on a particular complex of experimental conditions - a context - which can be chosen in a variety of ways. 
This sensitivity called contextuality (situatedness, embeddedness) \cite{Khrennikov2003,Gabora2005,Khrennikov2009b,Surov2019a} is ubiquitous in human cognition and behavior \cite{Schwarz1992,Nardi1996,Schwarz1999,Schwarz2007,Aerts2000a}. 


\paragraph{The qubit}

In relation to a given objective uncertainty, each possible context is represented as a vector in the complex-valued linear vector (Hilbert) space basis of which consists of possible behavioral alternatives.
In case of an objective uncertainty with two potential outcomes $0$ and $1$ each context is represented by a two-dimensional vector
\begin{equation}
\label{eq1}
\ket{\mathrm\Psi} = c_0\ket{0} + c_1\ket{1}, \quad \left|c_{0}\right|^2 + \left|c_{1}\right|^2 = 1
\end{equation}
called qubit where <<kets>> $\ket{i}$ denote the orthogonal basis vectors $(\braket{0}{1}=0)$ and $c_i$ are complex-valued amplitudes of the corresponding alternatives. Any normalized linear combination of vectors \eqref{eq1} is another vector in the same Hilbert space, possibly representing some other context of the considered uncertainty.

\begin{figure*}[t]
	\centering
	\includegraphics[width=0.75\textwidth]{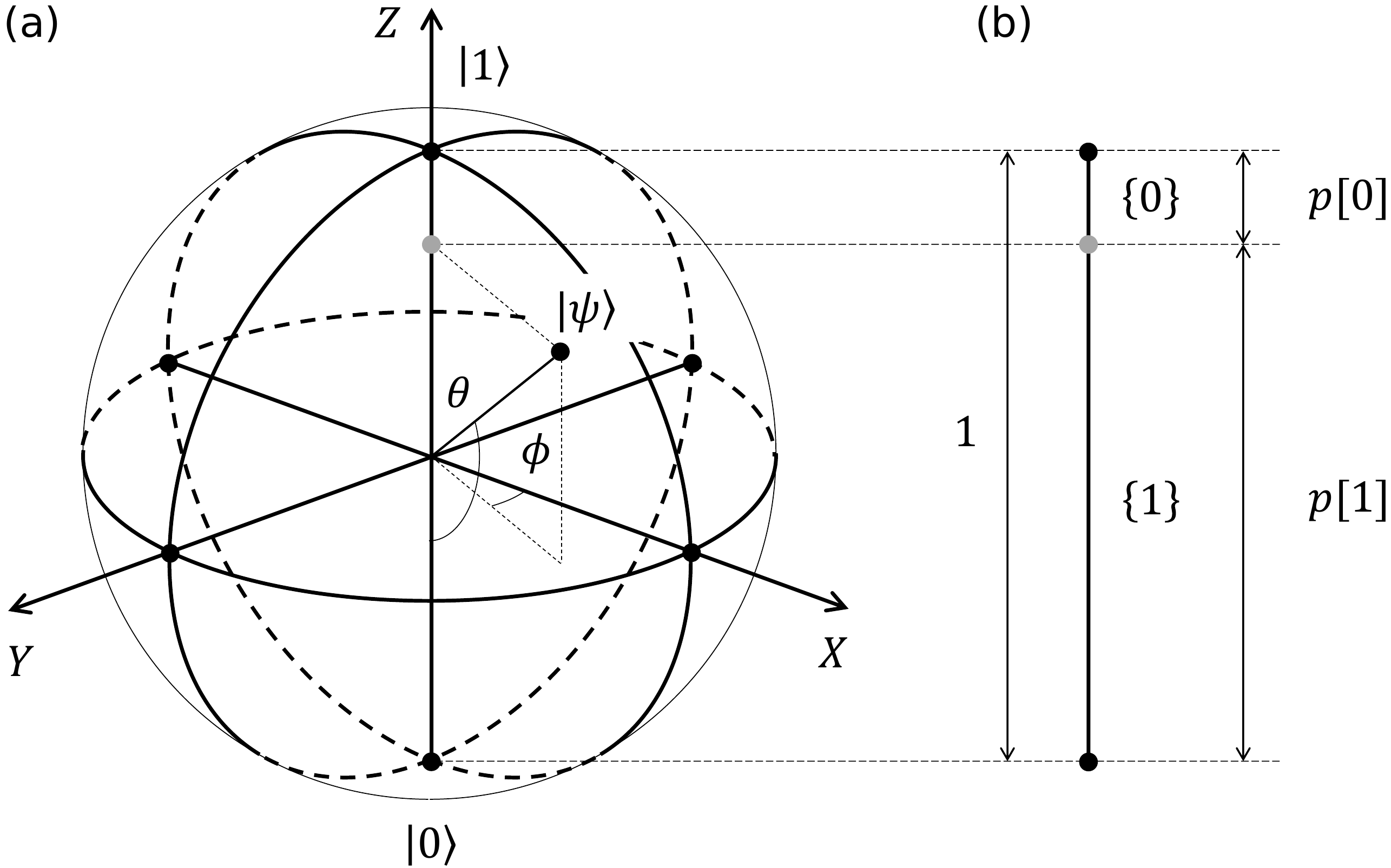}
	\caption{(a) The Bloch sphere. Cognitive vector state $\ket{\mathrm\Psi}$ \eqref{eq4} represents a particular behavioral context in relation to the objective uncertainty with alternatives 0 and 1. Polar angle $\theta$ defines probabilities of the possible outcomes 0 and 1 \eqref{eq5} if the experiment would be performed. Azimuthal phase $\phi$ encodes relation with other cognitive representations and has no mechanical analogy. (b) Classical probability space of subjective uncertainty possibly remaining after actualization of the objective uncertainty (a). Alternative events $\{0,1\}$ are represented by parts of a unit segment corresponding to the diameter of the Bloch sphere on the left.}
	\label{fig1}
\end{figure*}

In the context represented by vector \eqref{eq1} the considered objective uncertainty resolves to outcomes $0$ and $1$ with probabilities \cite{Jaeger2007}
\begin{equation}
\label{eq2}
\nonumber
p[i] =\left| \braket{i}{\mathrm\Psi} \right|^2 = \left|c_{i}\right|^2, \quad i=0,1.
\end{equation}
Unity sum of probabilities \eqref{eq2} implies that vector \eqref{eq1} can be parametrized as
\begin{equation}
\label{eq4}
\ket{\mathrm\Psi} = e^{i \Phi} \left(\cos \dfrac{\theta}{2} \ket{0} + e^{i \phi} \sin \dfrac{\theta}{2}  \ket{1}\right),
\end{equation}
where polar angle $\theta \in [0,\pi]$ and azimuthal phase $\phi \in [0,2\pi]$ are 
\begin{equation}
\label{eq5}
\begin{aligned}
&\cos \dfrac{\theta}{2} = \left|c_0\right| = \sqrt{p[0]} \\
&\sin \dfrac{\theta}{2} = \left|c_1\right| = \sqrt{p[1]},  \qquad 
\quad e^{i\phi}=\dfrac{c_1}{c_0}
\end{aligned}
\end{equation}
and $\Phi$ is a global phase factor.
Setting the latter aside, vector \eqref{eq4} is represented by a unique point on a sphere defined by polar angles $\theta$ and $\phi$. This sphere called Bloch sphere is shown in Figure~\ref{fig1}(a).

Azimuthal phase $\phi$ describes the rotation of vector \eqref{eq4} around Z-axis thus having no direct connection with observable probabilities \eqref{eq5}. Instead, $\phi$ describes how context represented by $\ket{\mathrm\Psi}$ associates with other contexts relative to the considered objective uncertainty. This property of the azimuthal phase is central to the model developed in Sect.~\ref{model}.

Linear algebra of vectors \eqref{eq1} in the two-dimensional Hilbert space is the algebra of infinitely many possible contexts in which the considered objective uncertainty can be resolved. Except for its complex-valued structure discussed below, this algebra is the simplest possible option for vector representation of contexts.

\paragraph{Actualization of potentiality}

When an objective uncertainty resolves in an actual experiment, the resulting outcome is not necessarily known to a subject as described in Sect.~\ref{uncertainty}. This happens, for another example, when the quantum physical experiment finishes whereas the instrument's readings are unknown to the experimenter. In this case, the experiment selects one of the objectively multivariant potential futures as a single actual state of the system, whereas the uncertainty of a subject changes from objective to subjective type.

In the case of binary uncertainty, this actualization process allows for geometric visualization. Namely, vector $\ket{\mathrm\Psi}$ representing the experimental context projects to the diameter of the Bloch sphere splitting it to a pair of segments as shown in  Figure~\ref{fig1} (a) \cite{Aerts1995}. If the diameter is chosen to have unit length, the obtained segments are equal to probabilities \eqref{eq5} representing ignorance of the subject about the experimental outcome, panel (b).

\subsection{Cognitive state}
\label{cognitiveq}

Experimental practice shows that for the resolution of the elementary objective uncertainties studied in quantum physics, most of the environmental factors constituting experimental contexts (including actual positions and velocities of the elementary particles and configuration of the fields composing the setup) are not statistically important. 
In the case of binary uncertainties, all information relevant for probabilistic behavioral modeling can be aggregated in only two dimensions accounted by spherical coordinates $\theta$ and $\phi$ parametrizing the Bloch sphere shown in Figure~\ref{fig1}(a). In the Stern-Gerlach experiment, for example, these dimensions describe an orientation of the apparatus relative to the electron \cite{Feynman1964}.

This study explores a conjecture that the same holds for the behavior of the living organisms and humans in particular so that contexts of a macroscopic decision resolving an objective binary uncertainty can be mapped to the two-dimensional Hilbert space described in Sect.~\ref{context}.
This space in which contexts are accommodated and distinguished essentially refers to the notions of cognitive, conceptual, and semantic space of a subject \cite{Newby2001,Bruza2008b,Gardenfors2014}; the difference is that now this space is defined not absolutely but relative to the considered objective uncertainty. Representations of the particular contexts in this cognitive Hilbert space, e.g. vectors \eqref{eq4} in the case of the binary uncertainty 0/1, can be called cognitive states.


\section{Model: cognitive triad}
\label{model}

\subsection{Setup: dichotomic uncertainty in three contexts}
\label{setup}

Consider a subject (the black box of Sect.~\ref{bbb}) which may resolve an objective dichotomic uncertainty with the alternatives $0$ and $1$ (Sect.~\ref{uncertainty}) in three different contexts $a$, $b$ and $c$ (Sect.~\ref{context}).
The contexts $a$ and $b$ are defined by mutually exclusive states of some binary factor (yes/no, true/false, white/black, etc.) and represented by the subject in qubit cognitive Hilbert space (Sect~\ref{cognitiveq}) by vectors of the form \eqref{eq4}
\begin{equation}
\label{eq6}
\begin{aligned}
&\ket{\mathrm\Psi_a} = \cos \dfrac{\theta_a}{2} \ket{0} + \sin \dfrac{\theta_a}{2}  \ket{1}, \\
&\ket{\mathrm\Psi_b} = \cos \dfrac{\theta_b}{2} \ket{0} + e^{i \phi_{b}} \sin \dfrac{\theta_b}{2}  \ket{1},
\end{aligned}
\end{equation}
where zero azimuth is identified with the context $a$ so that $\phi_{a}=0$ and $\phi_{b}$ is azimuthal phase of state $\ket{\mathrm\Psi_b}$ relative to $\ket{\mathrm\Psi_a}$.

Polar angles $\theta_a$ and $\theta_b$ are related to the probabilities $p_a$ and $p_b$ with which the subject resolves the considered uncertainty, i.e. takes decisions $0$ and $1$ in the contexts $a$ and $b$ as prescribed by \eqref{eq5}
\begin{equation}
\label{eq7}
\begin{aligned}
&p_a[0] = \cos^2\dfrac{\theta_a}{2}, \qquad p_a[1] = 1-p_a[0] = \sin^2\dfrac{\theta_a}{2}, \\
&p_b[0] = \cos^2\dfrac{\theta_b}{2}, \qquad p_b[1] = 1-p_b[0] = \sin^2\dfrac{\theta_b}{2}.
\end{aligned}
\end{equation}

In the third context $c$ the subject is uninformed about the situation factor defining contexts $a$ and $b$ considering the corresponding states of the factor as equiprobable. Context $c$ is subjectively reflected to the cognitive state 
\begin{equation}
\label{eq8}
\begin{aligned}
\ket{\mathrm\Psi_c} =N \left(\ket{\mathrm\Psi_a}+e^{ix}\ket{\mathrm\Psi_b}\right)
= \cos \dfrac{\theta_c}{2} \ket{0} + e^{i \phi_c} \sin \dfrac{\theta_c}{2}  \ket{1},
\end{aligned}
\end{equation}
which is a superposition of states \eqref{eq6} discriminated by a phase factor $e^{ix}$ and normalized by a positive constant $N$. 
Cognitive state \eqref{eq8} generates decision probabilities
\begin{equation}
\label{eq10}
\begin{aligned}
p_c[0] & = N^2\left|\cos{\frac{\theta_a}{2}}+e^{ix}\cos{\frac{\theta_b}{2}}\right|^2 = \\ & = N^2 \left(p_a[0]+p_b[0] + 2\sqrt{p_a[0] p_b[0]} \cos x\right), \\ 
p_c[1] & = N^2 \left|\sin{\frac{\theta_a}{2}}+e^{i\left(\phi_{b}+x\right)}\sin{\frac{\theta_b}{2}}\right|^2 = \\
& =N^2 \left(p_a[1]+p_b[1] + 2 \sqrt{p_a[1] p_b[1]} \cos (\phi_{b} + x)\right),
\end{aligned}
\end{equation}
unit sum of which amounts to normalization of state \eqref{eq8}
\begin{equation}
\nonumber
\begin{aligned}
1&=\braket{\mathrm\Psi_c}{\mathrm\Psi_c} = p_c[0] + p_c[1] = \\
& =2N^2\left(1+\sqrt{p_a[0]p_b[0]} \cos x + \sqrt{p_a[1]p_b[1]} \cos{\left(\phi_{b}+x\right)} \right).
\end{aligned}
\end{equation}
Equations \eqref{eq10} have two distinct solutions
\begin{equation}
\label{eq9}
\begin{aligned}
x  = & \arccos \left[\dfrac{p_c[0] / N^2 - p_a[0]-p_b[0]}{2\sqrt{p_a[0] p_b[0]}}\right], \\ \phi_b = -x \pm & \arccos \left[\dfrac{p_c[1] / N^2 -p_a[1]-p_b[1]}{2\sqrt{p_a[1] p_b[1]}}\right].
\end{aligned}
\end{equation}
Through definitions \eqref{eq6} and \eqref{eq8} solution \eqref{eq9} with the positive and negative sign produces triples of the cognitive states $\ket{\mathrm\Psi_a}$, $\ket{\mathrm\Psi_b}$ and $\ket{\mathrm\Psi_c}$ of different properties.

\subsection{Degenerate and non-degenerate triads}
\label{triad}

Consider a hypothetical behavior such that probabilities of taking the decision $0$ in all three contexts are equal:
\begin{equation}
\label{eqID}
p_a[1]=p_b[1]=p_c[1]=p.
\end{equation}
Solutions \eqref{eq9} then reduce to
\begin{subequations}
\label{sol}
\begin{align}
\label{eq18}
& \phi_{b} =0  \ \ \rightarrow \ \ \phi_{c} =0 \\[-5pt]
\nonumber
x = \mathrm{arccos} \left[\frac{1}{2N^2}-1\right] \qquad \mathrm{and} \qquad \ & \\[-5pt]
\label{eq19}
& \phi_{b} =-2x \ \ \rightarrow \ \ \phi_{c} =-x
\end{align}
\end{subequations}
where \eqref{eq18} and \eqref{eq19} correspond to the positive and negative sign in \eqref{eq9} respectively.

According to \eqref{eq8} solution \eqref{eq18} implies that the contexts $a$, $b$ and $c$ are all mapped to a single cognitive state
\begin{equation}
	\label{eq181}
	\ket{\mathrm\Psi_a}=\ket{\mathrm\Psi_b}=\ket{\mathrm\Psi_c} = \sqrt{1-p} \ket{0}+ \sqrt{p} \ket{1},
\end{equation}
due to which solution \eqref{eq9} with the positive sign is called \textbf{degenerate}.

Solution \eqref{eq19}, in contrast, maps the contexts $a$, $b$, and $c$ to the cognitive states with distinct azimuthal phases
\begin{subequations}
\label{eq170}
\begin{align}
\label{eq171}
&\ket{\mathrm\Psi_a} = \sqrt{1-p}\ket{0} + \sqrt{p} \ket{1}\\
\label{eq172}
&\ket{\mathrm\Psi_b} = \sqrt{1-p}\ket{0} + e^{-2ix} \sqrt{p} \ket{1} \\
\label{eq173}
&\ket{\mathrm\Psi_c} = \sqrt{1-p}\ket{0} + e^{-ix} \sqrt{p} \ket{1}
\end{align}
\end{subequations}
even though they generate equal decision probabilities \eqref{eqID}.
Such \textbf{non-degenerate} triple of states produced by solution \eqref{eq9} with the negative sign is called \textbf{the cognitive triad}.

Each of vectors in the cognitive triad \eqref{eq170} can be represented in the Bloch sphere (Figure~\ref{fig1}) so that azimuthal phases $\phi_a=0$, $\phi_b$ and $\phi_c$ \eqref{eq19} define their orientation in the (horizontal) azimuthal plane XY.
This is shown in Figure~\ref{figNcirc} for different values of the normalization constant $1/2\leq N < \infty$ limited by domain of the $\arccos$ function in \eqref{eq19}.
 
Starting with $N=1/2$ corresponding to $x=0$ and degenerate triad \eqref{eq181}, increase of $N$ leads to a clockwise rotation of state $\ket{\mathrm\Psi_b}$ whereas $\ket{\mathrm\Psi_c}$ is oriented halfway between $\ket{\mathrm\Psi_a}$ and $\ket{\mathrm\Psi_b}$. 
Setting $N=1/\sqrt{2}$ results in $x=\pi/2$ which produces rectangular structure in which $\ket{\mathrm\Psi_a}$ and $\ket{\mathrm\Psi_b}$ oppose each other in the azimuthal plane while $\ket{\mathrm\Psi_c}$ lies in the YZ plane as shown in the second panel of Figure~\ref{figNcirc}.
The third panel corresponding to $N=1$ shows symmetrical triad where $x = 2\pi/3$.
The circle is completed in the limit $N\rightarrow \infty$ where $\ket{\mathrm\Psi_c}$ opposes coinciding states $\ket{\mathrm\Psi_a}$ and $\ket{\mathrm\Psi_b}$.

\subsection{Discrimination of contexts}
\label{discrim}

\begin{figure}
	\includegraphics[width=\textwidth]{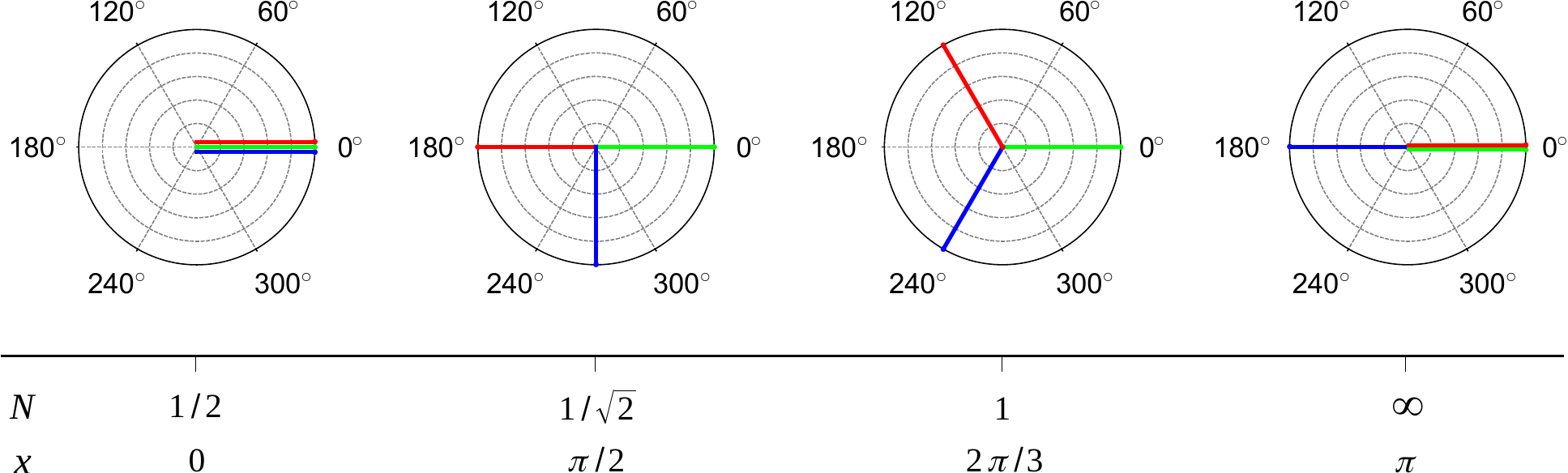}
	\caption{
	Cognitive triad \eqref{eq170} representing contexts $a$, $b$ and $c$ in projection to the azimuthal (XY) plane of the Bloch sphere (Figure~\ref{fig1}) for different normalization constants $N$ \eqref{eq8}.
	\textcolor{green}{$\ket{\mathrm\Psi_a}$}, \textcolor{red}{$\ket{\mathrm\Psi_b}$} and \textcolor{blue}{$\ket{\mathrm\Psi_c}$} are shown in \textcolor{green}{green}, \textcolor{red}{red} and \textcolor{blue}{blue} respectively. 
	Left: degenerate representation $N=1/2,x=0$ corresponding to rational logic \eqref{rational} and zero contextuality. $N=1/\sqrt2,x=\pi/2$: rectangular structure with maximal discrimination between the certainty contexts $a$ and $b$. $N=1$: symmetrical triad with $x=2\pi/3$ and uniform discrimination between all three contexts. Right: the limit $N\rightarrow\infty, x\rightarrow\pi$ of irrational logic with maximal discrimination of the uncertainty context $c$.}
	\label{figNcirc}
\end{figure}


Unlike degenerate solution \eqref{eq181}, cognitive triad \eqref{eq170} distinguishes contexts $a$, $b$, and $c$ even when they produce equal behavioral probabilities.
Following definition \eqref{eq8}, this discrimination is realized by the factors of type $e^{ix}$ defining phase delay between cognitive states in parallel with experimental observation \cite{TenOever2020}.
The resulting arrangement of cognitive states $\ket{\mathrm\Psi_a}$, $\ket{\mathrm\Psi_b}$, $\ket{\mathrm\Psi_c}$ in the azimuthal plane of the Bloch sphere depends on the normalization constant $N$ as shown in Figure~\ref{figNcirc}.


Variable discrimination between contexts $a$, $b$, and $c$ shown in different panels of Figure~\ref{figNcirc} can account for different subjective importance of an environmental factor defining these contexts in relation to the considered decision.
For example, charge but not color of a battery is important for a decision to use it or not. In this situation the contexts <<charged>> ($a$) and <<uncharged>> ($b$) ask for maximal discrimination as achieved by $N=1/\sqrt{2}$, while contexts <<white>>, <<black>> and <<unknown color>> (of the same battery) need not be distinguished at all, which corresponds to degenerate case $N=1/2$ (left panel in Figure~\ref{figNcirc}). For a decision concerned not by electrical performance but by visual appearance of the battery this preference reverses.

The uncertainty of the contextual factor can be subjectively more important then difference between definite states of the factor, as e.g. in case of strong ambiguity aversion associated with irrationality of the subject \cite{Khaw2017,Gorder2015}. The context <<unknown charge>> ($c$) then has to be discriminated better than in the symmetrical arrangement shown in the third panel of Figure~\ref{figNcirc}. This is realized by setting $N>1$. 

\paragraph{Energetic cost of cognitive discrimination}

Function of the constant $N$ is normalization of vector \eqref{eq8} where it defines how much cognitive states $\ket{\mathrm\Psi_a}$ and $e^{ix}\ket{\mathrm\Psi_b}$ have to be amplified or suppressed to combine to the normalized representation $\ket{\mathrm\Psi_c}$.
If one assumes that a neurophysiological implementation of this amplification requires energy \cite{Busemeyer2017a,Khrennikov2018a,Khrennikov2020}, then subjective ability to discriminate between the contexts $a$, $b$, $c$ defined by $N$ (Figure~\ref{figNcirc}) acquires real energetic cost measured in joules.

Degenerate cognitive representation produced by $N=1/2$ with no discrimination implies twofold suppression of $\ket{\mathrm\Psi_a}$ and $\ket{\mathrm\Psi_b}$ in amplitude so that most of the energy stored in these states can be released; in less extent, the same holds for rectangular structure generated by $N=1/\sqrt{2}$. 
In the case $N=1$ states $\ket{\mathrm\Psi_a}$ and $\ket{\mathrm\Psi_b}$ combine in full amplitudes without any suppression or amplification so that uniform discrimination between contexts $a$, $b$, $c$ by symmetrical cognitive triad is realized with zero energy balance.
Enhanced discrimination of the uncertainty context $c$ corresponding to  irrational logic is achieved by amplification of the cognitive amplitudes by $N>1$ which is energy-consuming.

\subsection{Azimuthal phase stability}
\label{stability}

Important feature of the cognitive triad (middle panels in Fig.~\ref{figNcirc}) is its ability to account for variable behavioral probabilities based on the stable structure of cognitive representations.
This property becomes apparent in comparison with degenerate representation \eqref{eq181} produced by \eqref{eq9} with the plus sign.

\begin{figure*}[t]
	\centering
	\includegraphics[width=\textwidth]{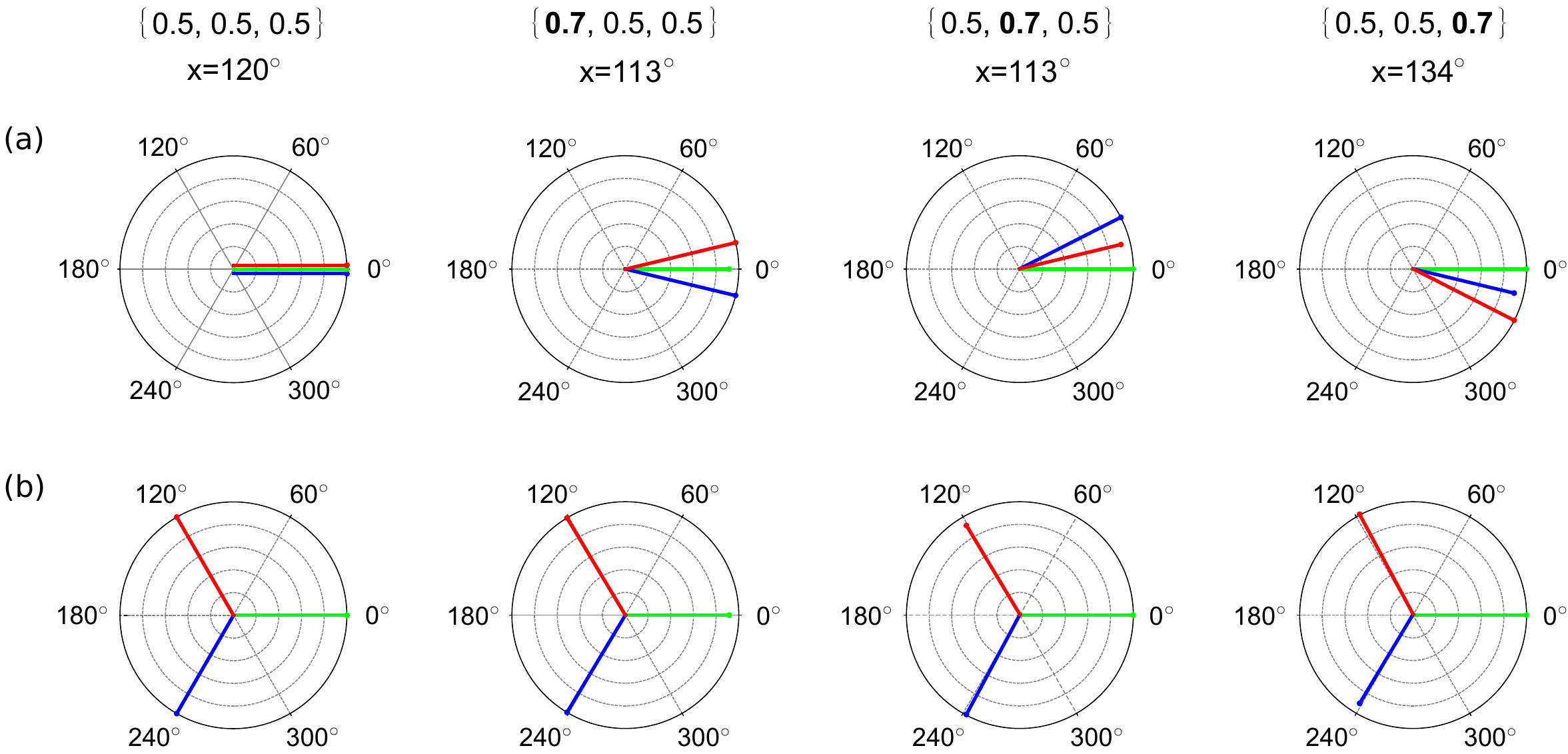}
	\caption{Azimuthal plane projections of the cognitive states \textcolor{green}{$\ket{\mathrm\Psi_a}$} \textcolor{green}{(green)}, \textcolor{red}{$\ket{\mathrm\Psi_b}$} \textcolor{red}{(red)} and \textcolor{blue}{$\ket{\mathrm\Psi_c}$} \textcolor{blue}{(blue)} representing the contexts $a$, $b$, $c$ which produce decision probabilities \textcolor{green}{$\{p_a[1]$}, \textcolor{red}{$p_b[1]$}, \textcolor{blue}{$p_c[1]\}$} indicated on top. Left: undisturbed case \eqref{eqID} same as panels 1 and 3 in Figure~\ref{figNcirc}. 
	(\textbf{a}) Degenerate solution ($N=1/2$ and plus sign in \eqref{eq9}) is strongly perturbed by variation of probabilities \eqref{eqVAR}, $\mathrm{\Delta}=0.2$.
	(\textbf{b}) In symmetrical triad ($N=1$ and minus sign in \eqref{eq9}) the same variation of probabilities is realized by tuning of only combinational phase $x$ (which is the same for both cases) while disturbance of the cognitive states is much smaller. This is mechanism of behavioral variability based on the stable structure of cognitive representations.}
	\label{figTrian}
\end{figure*}

Assume that one of identical probabilities \eqref{eqID} is disturbed as 
\begin{equation}
\label{eqVAR}
p_i[1]\rightarrow p+\mathrm{\Delta}
\end{equation}
so that solutions considered in Sect.~\ref{triad} no longer hold; in this case degeneracy \eqref{eq181} is lifted by amount of the first order in $\mathrm{\Delta}$ as shown in Figure~\ref{figTrian}(a) for $\mathrm{\Delta}=0.2$.

In contrast, the same disturbance of decision probabilities causes almost no change in the representation of contexts by cognitive triad \eqref{eq170} as shown in Figure~\ref{figTrian}(b). Modification of behavioral probabilities in this case is realized by tuning exclusively the combinational phase $x$ which is identical in both solutions \eqref{eq9}. 
Based on experimental modeling reported in Sect.~\ref{experiment} this property is further discussed in Sect.~\ref{semstability}.


\subsection{Context Representation Algorithm}
\label{algorithm}

Given experimentally measured probabilities $p_a$, $p_b$, $p_c$, building representation of the contexts $a$, $b$, and $c$ according to the model described in Sect.~\ref{model} consists in following steps:
\begin{enumerate}
	\item polar angles $\theta_a$ and $\theta_b$ are determined from \eqref{eq7};
	\item the normalization constant $N$ defining arrangement of vectors in the cognitive triad (Figure~\ref{figNcirc}) is chosen according to the importance of the situational factor defining contexts $a$, $b$, and $c$ (Sect.~\ref{discrim})\footnote{Generally two different constants allowing for asymmetric composition of $\ket{\mathrm\Psi_a}$ and $\ket{\mathrm\Psi_b}$ in \eqref{eq8}.};
	\item phases $x$ and $\phi_{b}$ are calculated according to \eqref{eq9} with the minus sign corresponding to non-degenerate solution. With polar angles found in step 1 this defines cognitive states $\ket{\mathrm\Psi_a}$ and $\ket{\mathrm\Psi_b}$ \eqref{eq6};
	\item cognitive state $\ket{\mathrm\Psi_c}$ is calculated from \eqref{eq8} with polar angle $\theta_c$ and azimuthal phase $\phi_c$ determined according to \eqref{eq5}.
\end{enumerate}

The resulting representation consisting of three qubit cognitive states $\ket{\mathrm\Psi_a}$, $\ket{\mathrm\Psi_b}$, and $\ket{\mathrm\Psi_c}$ that can be visualized by points on the Bloch sphere (Figure~\ref{fig1}).
This output is encoded by 5 parameters, namely
polar angles $\theta_a$, $\theta_b$, $\theta_c$ and azimuthal phases $\phi_b$, $\phi_c$.


\paragraph{Existence of representation}

\begin{figure*}[t]
	\centering
	\includegraphics[width=\textwidth]{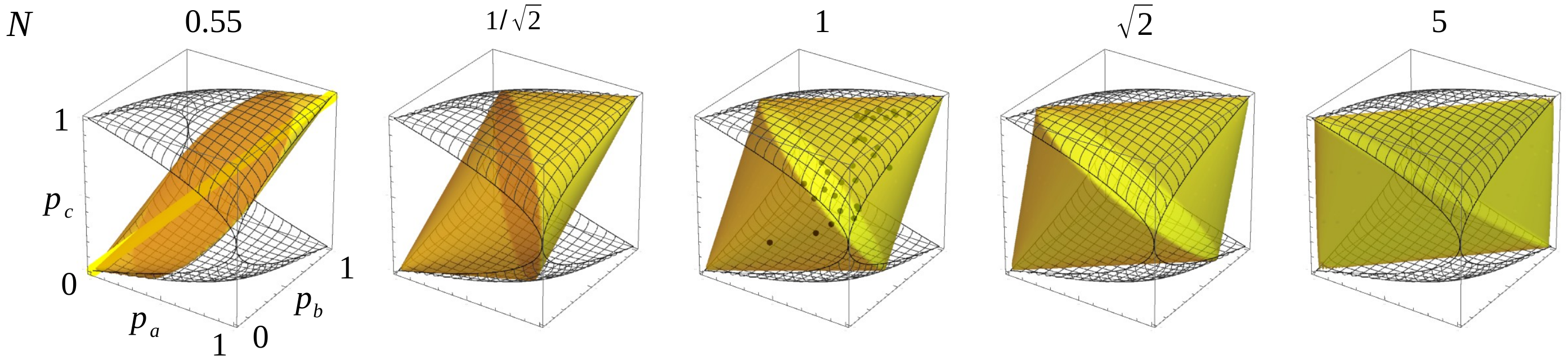}
	\caption{Regions in probability space $p_a[1]\times p_b[1]\times p_c[1]$ for which representation of the contexts $a$, $b$ and $c$ exists for different normalization constants $N$.
	\textbf{Left} panel, $N=0.55$: thin leaf restricting probabilities by rational logic relation \eqref{rational} reducing to the diagonal \eqref{eqID} in degenerate case $N=1/2$.
	\textbf{Center}, $N=1$: symmetric bicone corresponding to symmetric triad. All 32 experimental probability triples (Sect.~\ref{experiment}) shown by dots are in the allowed region.
	\textbf{Right}, $N=5$: thin vertical slab along diagonal of the cube confining probabilities to the maximum irrationality case \eqref{irrational}.
	Meshed surfaces show boundaries \eqref{eq12} within which context's representation exists for some $N$.} 
	\label{figPS}
\end{figure*}

Step 3 in the above algorithm is possible if and only if arguments of the $\arccos$ functions in \eqref{eq9} do not exceed unity by modulus.
Probability triples $p_a$, $p_b$, $p_c$ which satisfy these conditions for different $N\geq1/2$ are visualized in the Cartesian probability cube $p_a\times p_b\times p_c$ as shown in Figure~\ref{figPS}. 

In degenerate case $N=1/2$ solution exists only in trivial case \eqref{eqID} which is diagonal of the cube shown in the left panel.
At $N=0.55$ this diagonal is expanded to a leaf-like shape, same panel.
Within the width of the leaf probabilities obey relation 
\begin{equation}
\label{rational}
p_c \approx \dfrac{p_a+p_b}{2}
\end{equation}
which is associated with rational logic \cite{Khrennikov2015c}.
As $N$ increases, the allowed probability region gets thicker and transforms to a bicone structure so that rational relation \eqref{rational} no longer holds; at $N=1$ the bicone gets symmetric and occupies a maximal fraction of nearly 70 percent of the probability cube.
Further increase of $N$ squeezes the bicone which in the limit $N \rightarrow \infty$  reduces to the plane 
\begin{equation}
\label{irrational}
p_a \approx p_b, \quad p_c \in \left[0,1\right]
\end{equation}
shown in the right panel. This is the limit of maximal irrationality when opposite states of the context-defining factor produce nearly the same decision probabilities $p_a \approx p_b$ while in the uncertainty context $c$ probability can take all possible values.

Representation of the contexts $a$, $b$, $c$ according to the above algorithm exists if equations \eqref{eq9} are solvable in real phases $\phi_b$ and $x$ at least for some $N$. This is possible if and only if
\begin{equation}
\label{eq12}
\dfrac{p_a+p_b-2\sqrt{p_ap_b}}{1+\sqrt{(1-p_a)(1-p_b)}-\sqrt{p_ap_b}} \leq 2p_c \leq 
\dfrac{p_a+p_b+2\sqrt{p_ap_b}}{1-\sqrt{(1-p_a)(1-p_b)}+\sqrt{p_ap_b}},
\end{equation}
where all probabilities refer to the outcome 1 so that $p_i=p_i[1]$ in the notation used above.
Upper and lower boundaries for $p_c$ \eqref{eq12} are shown in Figure~\ref{figPS} by meshed surfaces. The resulting domain constitutes a union of allowed probability regions corresponding to all possible values of normalization constant $1/2 \leq N <\infty$ exemplified in Figure~\ref{figPS}.
Domain \eqref{eq12} is symmetric relative to the plane $p_a=p_b$ at which solution is possible for any $0\leq p_c \leq 1$ \eqref{irrational}. Most restricted cases are $p_a=1$, $p_b=0$ and vise versa where the allowed range for $p_c$ shrinks to a single point $p_c=0.5$.

\paragraph{Relation to QLRA}

The representation scheme just described is related to a quantum-like representation algorithm (QLRA) which models the probabilistic logic of two dichotomic decisions in a single two-dimensional Hilbert space \cite{Khrennikov2008a}.
With the resolution of one dichotomic uncertainty considered as a context for subsequent resolution of the remaining one, correspondence with the setup considered in this paper is established. 

The difference between the two representations is that QLRA maps exclusive contexts ($a$ and $b$ in this paper) to orthogonal quantum states, while in the representation reported here this restriction is lifted. As explained in Sect.~\ref{discrim}, this accounts for the possibility that difference between the represented contexts for the considered decision may have varying subjective importance: when such importance is low then these contexts, even though mutually exclusive, need not be discriminated well and therefore can be represented by close cognitive states.

\section{Experiment: the two-stage gamble}
\label{experiment}

\subsection{Data}
\label{data}

General behavioral structure accounted by the model developed in Sect.~\ref{model} is realized in a so-called two-stage gambling experiment originally devised to reveal the irrationality of human decision making \cite{Tversky1992}.
By design, subjects are exposed to an alternative to play or not to play in a gamble in three different contexts, namely when the previous round is won, lost, or has an unknown outcome with 50/50 probability.
Each round of the game consists in guessing the results of a fair coin tossing. Successful guess is rewarded with 2 monetary units while the mistake is penalized by 1 monetary unit.

\begin{table}[h]
	\caption{Experimental data on the two-stage gambling task. The rows correspond to the measured probabilities to play in the next round of the gamble in three contexts when the previous round is won ($a$), lost ($b$), and unknown with 50/50 chance ($c$). Experiments 1-4: ref. \cite{Tversky1992}; 5-8: ref. \cite{Kuhberger2001}; 9-11: ref. \cite{Lambdin2007}; 12: ref. \cite{Surov2019}; 13-32: ref. \cite{Broekaert2020}. In the latter group, experiments 13-17 correspond to the <<between subjects>> setup; 18-22: <<within subjects>> setup (WS) with random order of contexts; 23-27: WS with <<Known>>$\rightarrow$<<Unknown>> order of contexts; 28-32: WS with <<Unknown>>$\rightarrow$<<Known>> order of contexts. In each of the series 13-17, 18-22, 23-27, 28-32 experiments are arranged in ascending order of the payoff parameter.}
	\label{tab1}       
	\begin{tabular}{ccccccccccccc}
		\toprule\noalign{\smallskip}
		\thead{No.} & 1  & 2  & 3 & 4 & 5 & 6 & 7 & 8 & 9 & 10 & 11 & 12   \\
		\hline\noalign{\smallskip}
		\thead{<<Won>> \\ $p_a[1]$} &.69&.75&.69&.71&.60&.83&.80&.68&.64&.53&.73&.30\\
		\thead{<<Lost>> \\ $p_b[1]$} &.57&.69&.59&.56&.47&.70&.37&.32&.47&.38&.49&.24\\
		\thead{<<Unknown>> \\ $p_c[1]$} &.38&.73&.35&.84&.47&.62&.43&.38&.38&.24&.60&.17\\
		\toprule\noalign{\smallskip}
		\thead{No.} & 13  & 14  & 15 & 16 & 17 & 18 & 19 & 20 & 21 & 22 & &    \\
		\hline\noalign{\smallskip}
		\thead{<<Won>> \\ $p_a[1]$} &.82&.75&.65&.58&.56&.75&.72&.64&.57&.55&&\\
		\thead{<<Lost>> \\ $p_b[1]$} &.92&.89&.87&.85&.85&.86&.83&.81&.80&.77&&\\
		\thead{<<Unknown>> \\ $p_c[1]$} &.87&.86&.87&.85&.81&.89&.86&.84&.84&.85&&\\
		\noalign{\smallskip}\hline
		\toprule\noalign{\smallskip}
		\thead{No.} & 23  & 24  & 25 & 26 & 27 & 28 & 29 & 30 & 31 & 32 & \multicolumn{2}{c}{Mean}    \\
		\hline\noalign{\smallskip}
		\thead{<<Won>> \\ $p_a[1]$}        &.69&.61&.56&.45&.42&.68&.62&.52&.48&.43&\multicolumn{2}{l}{.625$\pm$.022} \\
		\thead{<<Lost>> \\ $p_b[1]$}         &.74&.65&.60&.59&.53&.66&.65&.51&.51&.41&\multicolumn{2}{l}{.629$\pm$.032} \\
		\thead{<<Unknown>> \\ $p_c[1]$} &.64&.59&.48&.41&.37&.78&.75&.58&.63&.52&\multicolumn{2}{l}{.620$\pm$.038} \\
		\noalign{\smallskip}\bottomrule
	\end{tabular}
\end{table}

In this setup decisions <<to play>> and <<not to play>> are mutually exclusive and complementary alternatives denoted $1$ and $0$ in the model above. 
Outcome of the previous round of the same game serves as a two-valued situation factor defining contexts $a$, $b$, $c$ as <<won>>, <<lost>>, and <<unknown>> respectively.
The result of the experiment is aggregated to a triple of statistical probabilities $p_a[1]$, $p_b[1]$, $p_c[1]$ calculated as the number of subjects who decided to play divided by the total number of respondents in each context. 

In total 32 experiments of this type reported in \cite{Tversky1992,Kuhberger2001,Lambdin2007,Surov2019,Broekaert2020} were used to test the representational algorithm described in Sect.~\ref{algorithm}. The raw probability data are shown in Table~\ref{tab1}.

\subsection{Modeling}
\label{results}

\begin{figure*}
	\includegraphics[width=\textwidth]{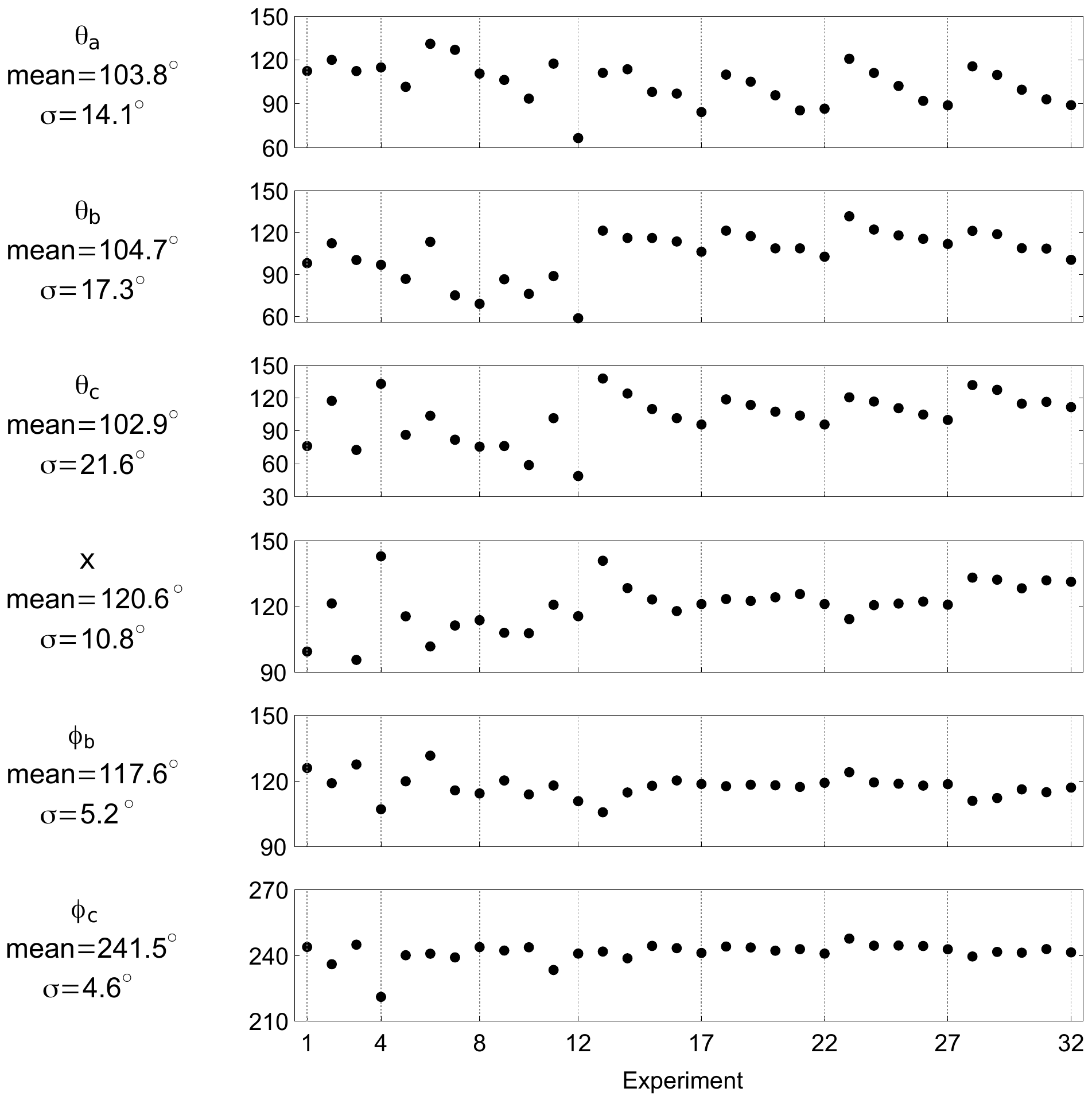}
	\caption{Parameters of the context representation models returned by algorithm of Sect~\ref{algorithm} for 32 probability triples listed in Table~\ref{tab1}. Polar angles $\theta_a$, $\theta_b$, $\theta_c$ define  probabilities to play or not to play the second round of the gamble in the contexts <<won>>, <<lost>> ans <<unknown>> according to \eqref{eq5}. Lower panels show combinational phase $x$ and azimuthal phases $\phi_{b}$, $\phi_c$ \eqref{eq8} encoding semantics of the contexts relative to the decision.}
	\label{fig2}
\end{figure*}

The context representational algorithm (Sect~\ref{algorithm}) is tested in the non-degenerate mode defined by the minus sign in \eqref{eq9} for the data aggregated from 32 experiments described in Sect.~\ref{data}.
Normalization constant $N$ was set to $1$ corresponding to the symmetrical cognitive triad mode (panel 3 in Figure.~\ref{figNcirc}).
With this choice, representation exists for all 32 probability triples listed in Table~\ref{tab1} and shown by dots in the center panel of Figure~\ref{figPS}.
Polar angles $\theta_a$, $\theta_b$, $\theta_c$ and azimuthal phases $\phi_{b}$, $\phi_c$ returned by the algorithm for each experiment are shown in Figure~\ref{fig2}.

\begin{figure*}[h]
	\centering
	\includegraphics[width=\textwidth]{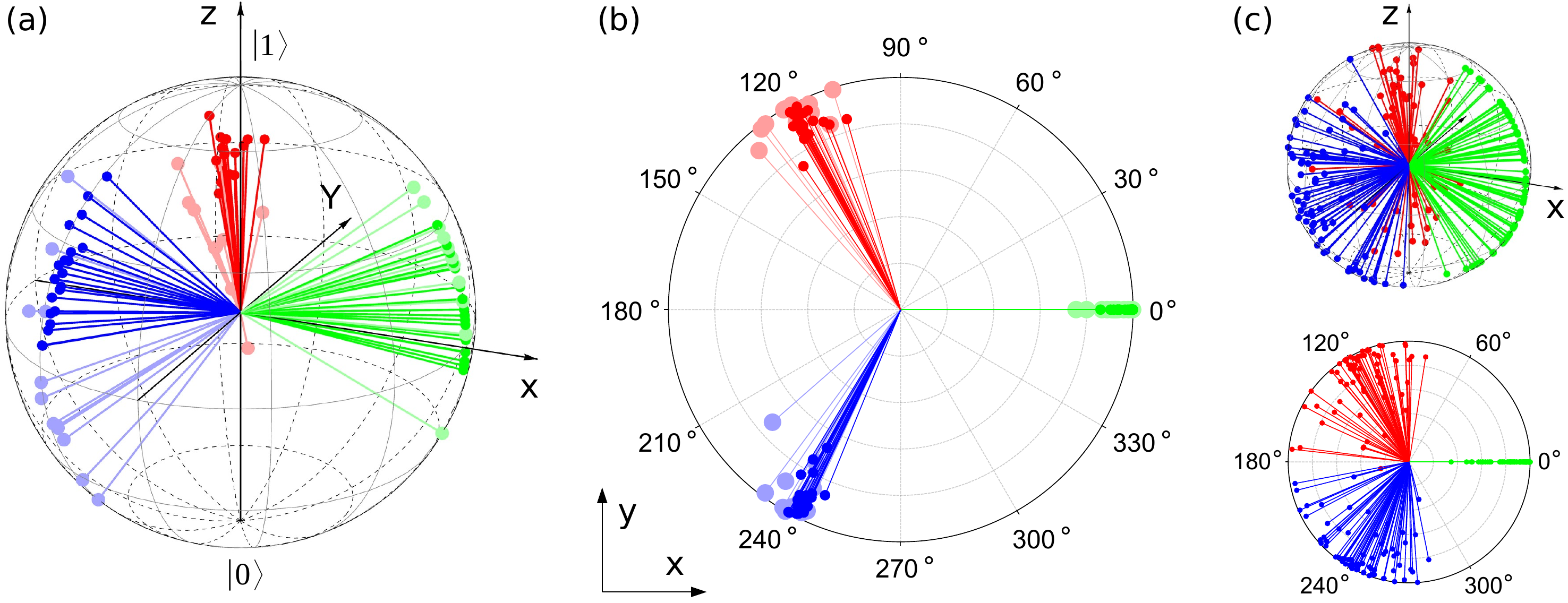}
	\caption{Cognitive triad models for each of the 32 two-stage gambling experiments in the Bloch sphere representation (Figure~\ref{fig1}) defined by parameters shown in Figure~\ref{fig2}. (a) 3D view of the cognitive states \textcolor{green}{$\ket{\mathrm\Psi}_a$ (<<won>> context $a$, green)}, \textcolor{red}{$\ket{\mathrm\Psi}_b$ (<<lost>> context $b$, red)}, and \textcolor{blue}{$\ket{\mathrm\Psi}_c$ (<<unknown>> context $c$, blue)} in the Bloch sphere. (b) Projection of the same states to the azimuthal (XY) plane where azimuthal phases $\phi_{b}$ and $\phi_c$ define orientation of $\ket{\mathrm\Psi}_b$ and $\ket{\mathrm\Psi}_c$ relative to $\ket{\mathrm\Psi}_a$ taken as zero. 
	(c): the same for randomly generated probability triples.}
	\label{fig3}
\end{figure*}


The mean values of polar angles and azimuthal phases shown on the left of Figure~\ref{fig2} define a single cognitive triad representing contexts $a$, $b$, $c$ averaged over all 32 experiments.
Since mean values of the polar angles are nearly equal, this single cognitive triad is closely approximated by a baseline solution \eqref{eq170} with $p \approx 0.62$ and combinational phase $x\approx120^\circ$ corresponding to the symmetric mode $N=1$.
This can be checked by observing that averaged probabilities $p_a$, $p_b$, $p_c$ indicated in the last column of Table~\ref{tab1} coincide up to statistical error thereby satisfying the identity condition \eqref{eqID}.

Parameters shown in the panels of Figure~\ref{fig2} define a triple of the qubit cognitive states $\ket{\mathrm\Psi_a}$, $\ket{\mathrm\Psi_b}$, $\ket{\mathrm\Psi_c}$ representing the contexts $a$ (green), $b$ (red), $c$ (blue) in each of 32 experiments. The Bloch sphere representation of these states is shown in panels (a) and (b) of Figure~\ref{fig3}. Panel (a) gives a full three-dimensional view. Panel (b) shows the projection of states to the azimuthal XY plane.

For comparison, the algorithm was applied in the same  symmetrical triad mode to synthetic data consisting of 100 randomly generated probability triples. 88 of them satisfied condition \eqref{eq12} and 71 found representation in the symmetrical triad mode with $N=1$ (i.e. fell within the bicone shown in the middle panel of Figure~\ref{figPS}).
The obtained 71 triads of the context representation states are shown in Figure~\ref{fig3}(c).

\subsection{Azimuthal phase stability}
\label{phaseReg}

Azimuthal phases $\phi_{b}$, $\phi_c$ and the combinational phase $x$ obtained from the context representation algorithm have distributions
\begin{subequations}
	\label{eqPhases}
	\begin{align}
	\label{eq15}
	&x = 120.6 \pm 10.8^\circ \\
	\label{eq14}
	&\phi_{b} = 117.6 \pm 5.2^\circ\\
	\label{eq16}
	&\phi_c = 241.5 \pm 4.6^\circ.
	\end{align}
\end{subequations}
As evident from Figure~\ref{fig3}, standard deviations \eqref{eq14} and \eqref{eq16} are radically lower than obtained for probability triples generated randomly; they are also 3 to 4 times lower than standard deviation of polar angles $\theta_a$, $\theta_b$ and $\theta_c$ indicated in Figure~\ref{fig2}. 
Moreover, Figure~\ref{fig2} shows that azimuthal phases $\phi_{b}$ and $\phi_c$ have no significant trace of regular patterns exhibited by polar angles $\theta_a$, $\theta_b$ and $\theta_c$ in series of experiments 13-32. 
These observations indicate that near-constant behavior of $\phi_{b}$ and $\phi_c$ is neither imposed by the modeling scheme nor constitutes an incidental byproduct of overall homogeneity of the probability data. 

Invariance of the values $\phi_{b}$ and $\phi_c$ with respect to particularities of behavioral contexts in different realizations of the experiment is a non-trivial phenomenon; called quantum phase stability, it was previously observed via standard quantum model of the two-stage gamble \cite{Surov2019}.
Now, however, stability refers to a particular dimension in the structure of subjective context representations, namely azimuthal phase of cognitive states in the qubit Hilbert space.

\section{Semantics and behavior}
\label{semantics}

\subsection{Azimuthal phase as semantic dimension}
\label{azimuthsemantics}

Azimuthal degree of freedom allows a subject to distinguish behavioral contexts irrespective of the decision probabilities associated with them; as discussed in Sect.~\ref{discrim}, this discrimination can account for varying subjective importance of the factor defining these contexts, which in the two-stage gambling setup is an outcome of the previous round of the game. 
Accommodation of cognitive states in the azimuthal plane of the Bloch sphere allows a subject to recognize behavioral contexts according to this subjective importance.

The resulting cognitive organization reflects a meaningful part of contextual information - that part which is subjectively relevant for making the considered decision, i.e. resolving an objective uncertainty described in Sect.~\ref{uncertainty}.
In line with the existing approaches to semantics \cite{Newby2001,Gershenson2007,Bruza2008b,Gardenfors2014,Kolchinsky2018b,DeJesus2018,Galofaro2018}, the azimuthal phase dimension $\phi$ of the qubit cognitive states functions as one-dimensional personal semantic space involved in a particular behavioral task.

\subsection{Semantic stability and behavioral variability}
\label{semstability}

\paragraph{Specialization of phases}

According to Sect.~\ref{azimuthsemantics}, compactness of azimuthal phases \eqref{eq14} and \eqref{eq16} is interpreted as stability of subjective semantic relations between contexts $a$, $b$, $c$ involved in the decision to play or not in the second stage of the gamble. 
Compared to the initial observation \cite{Surov2019}, in the cognitive triad model reported here this phenomenon of quantum phase stability acquires an explicit mechanism.
Now, stability refers to azimuthal phases $\phi_{b}$ and $\phi_c$ specifying a fixed cognitive structure of a subject, while tuning of decision probabilities involves an additional degree of freedom - the combinational phase $x$. The latter defines the composition of stable cognitive representations of contexts \eqref{eq8} used to generate varying decision probabilities Table~\ref{tab1}.
The corresponding fluctuation of polar angles \eqref{eq5} is shown in the top panels of Figure~\ref{fig2}.

Specialization of the combinational phase $x$ and representational phases $\phi_{b}$, $\phi_c$ is a crucial feature of the cognitive triad model and the resulting algorithm for context representation (Sect.~\ref{algorithm}).
A subject who uses this cognitive algorithm can adjust decision probabilities according to the details of a particular situation based on a stable representation of gross behavioral contexts thereby combining behavioral variability and cognitive stability.
As required for cognition \cite{Atmanspacher2005}, this relieves an individual from a need to reconfigure his or her neuronal system each time when a particular decision has to be made. Instead, a single phase relation $x$ realized e.g. by a temporal delay between the corresponding neuronal oscillation modes \cite{Barrett1969,Plikynas2016,DeBarros2017} has to be tuned\footnote{In close analogy with an algorithm for sorting of contextual representations based on neural phase encoding \cite{TenOever2020}.}.
In this manner, a stable yet flexible cognitive model of individual activity, arguably favored by natural selection \cite{Gabora2009a,Krakauer2010}, can function.

\paragraph{Semantic stability as triadic feature}

Regularities extracted from the data by the model above are not encountered when decision probabilities $p_a$, $p_b$, $p_c$ (Table~\ref{tab1}) are analyzed in pairs. For example, stable relation between cognitive representations $\ket{\mathrm\Psi_a}$ and $\ket{\mathrm\Psi_c}$ defining decision probabilities in contexts $a$ and $c$ (Figure~\ref{fig3}) cannot be established from values $p_a$ and $p_c$ (Table~\ref{tab1}) which show practically no correlation with $R^2\approx0.18$.
An intuition for this is that once any context of the triad $a$, $b$, $c$ is ignored, then a semantic account of behavioral data becomes impossible; what remains is a non-contextual, objectified correlation which is an unsuitable basis for modeling of cognition \cite{Tafreshi2016}. The dropped element then acts as an uncontrolled confounder spoiling the observed binary correlation \cite{Pearl2000}. 

Contrary to correlation, quantum phases are inherently triadic parameters capturing relations between behavioral probabilities undetectable at the dyadic level of analysis. The significance of the triadic structure in cognitive modeling is further discussed in Sect.~\ref{carrier}.

\subsection{Prognosis of behavior}
\label{prognoz}

Stability of the quantum phase parameters can be projected to other experiments including that which are not yet performed.
In the case of the two-stage gambling experiment, the quantum phase stability holds for subject groups sampled across decades and different countries, allowing for the probabilistic prognosis of behavior \cite{Surov2019}. Figure~\ref{fig2} shows that azimuthal phases found for experiments 13-32 which appeared since 2019 align with the earlier data thereby supporting this method of predictive behavioral modeling.

Prognosis of behavior based on the quantum phase stability parallels prediction of survey responses based on semantic analysis of texts and lexical databases \cite{Arnulf2018,Arnulf2020a}. Semantics, extracted in this approach from thematic corpora specific to the considered subject group, represents regularities of the corresponding collective cognition. In quantum modeling these regularities are considered as internal mechanics of the black box (Sect.~\ref{method}) reflected by stable phase parameters of quantum models.

\section{Outlook}
\label{outlook}

\subsection{Semantic mode of quantum modeling}
\label{smode}

\paragraph{Modeling of irrationality}

The two-stage gambling setup was previously considered in quantum modeling as behavioral case compromising probabilistic expectations of rational logic by so-called disjunction fallacy, violations of logical distributivity and a so-called sure thing principle \cite{Tversky1992,Khrennikov2015c}.
Later experiments indicated that such violations are not necessarily observed \cite{Lambdin2007}; as seen from Table~\ref{tab1} and Figure~\ref{fig2}, average probabilities $p_a$, $p_b$, $p_c$ and the corresponding polar angles are practically equal in agreement with rational (Boolean) logic \cite{Khrennikov2015c}.
Absence of irrationality in probability data, however, does not hinder the quantum representation of behavioral contexts as shown in Sect.~\ref{triad} where <<rational>> probability triple $p_a=p_b=p_c$ \eqref{eqID} is considered. 
This illustrates the use of quantum theory in a mode that is essentially different from the explanation of non-classical probability patterns.

\paragraph{Semantic mode}

Interpretation of azimuthal phases of complex-valued probability amplitudes as semantic dimensions (Sect.~\ref{azimuthsemantics}) renders quantum models of cognition and behavior as models of subjective meaning behind the observable processes. This mode of quantum modeling is developed e.g. in \cite{Melucci2005,Widdows2007,Bruza2008b,Gabora2017,Busemeyer2018,Aerts2018,Surov2019a}.

The method for representation of contexts described in this paper exercises this semantic mode of quantum behavioral modeling; in short, it aims to model subjectively contextual sense-making of behavior which is intrinsically human cognitive strategy \cite{Cosmelli2008,Salvatore2020}.
Quantum-theoretical modeling approach described above
\begin{enumerate}
\item considers cognitive representations as non-separable unitary states (Sect.~\ref{cognitiveq}); 
\item considers semantics as a subjective model for potential context-dependent behavior (Sect~\ref{uncertainty});
\item represents semantic relations in low-dimensional Hilbert space constructed for local decision making situation (Sect.~\ref{context});
\item allows for statistical prognosis of behavior based on the phase stability relations (Sect~\ref{semstability}, cf. \cite{Glockner2012}),
\end{enumerate}
thereby conforming with general properties of semantic models envisioned in \cite{Cosmelli2008,Samsonovich2009}.
Behaviorism, in its quantum version (Sect~\ref{method}), thus can be indeed considered as an instrument for a quantitative science of meaning \cite{Osgood1952,DeGrandpre2000}.

The quantum cognitive triad and the algorithm for context representation described in Sect.~\ref{model} realize features listed above in the simplest nontrivial case of dichotomic decision. 
The model is generalizable to multi-optional uncertainties and a larger number of contexts.
Supplemented with azimuthal phase regularities (Sect.~\ref{semantics}), it constitutes an approach for solving the quantum phase problem stated in Sect.~\ref{adventQuantum}.

\subsection{Quantification of subjectivity}
\label{reflexivity}


\paragraph{Subject and uncertainty}

Subjectivity understood as <<the influence of personal beliefs or feelings, rather than facts>> \cite{SubjectDef} implies the ability of an agent to act, that is, to resolve an objective uncertainty \cite{Popper1978a,Morf2018a}; potential contexts of this resolution are then represented in the personal cognitive Hilbert space constructed ad hoc (Sect.~\ref{context}). 

In the simplest setup modeled above, subjectivity of this mapping is accounted by azimuthal phase $\phi$ which contrary to polar angle $\theta$ in the Bloch sphere (Figure~\ref{fig1}(a)) is not predetermined by decision probability in the considered context, but is for a subject to decide\footnote{The Bloch sphere thus can be considered as a visualization tool for subjective semantics of certainty and uncertainty \cite{Hullman2020}.}. This non-uniqueness is a basis of subjective relativity which lies at the very core of quantum methodology of behavioral modeling.

The predetermined processes like the free-fall motion considered in Sect.~\ref{uncertainty}, in contrast, are <<based on facts>> and have no room for an agency to intervene. Personal ignorance about objective events of this kind is quantified by the division of the Bloch sphere's diameter as shown in Figure~\ref{fig1}. Contrary to the point on a surface, this division is uniquely defined by the probability of the measurement outcome and does not have an additional degree of freedom analogous to azimuthal phase $\phi$ which could capture subjectivity of an agent.

\paragraph{Quantitative human sciences}

Inconsistency between the subjective nature of human cognition and objective descriptions of the latter kind is considered as a fundamental restriction forbidding the development of quantitative models in psychology \cite{Tafreshi2016}, Sect.~\ref{clb}. 
The method of behavioral modeling developed in this paper, however, points to a way out of this deadlock by showing that identification of quantitativeness and objectivity is incorrect. In fact, subjectivity can be quantified in quantum model of semantic representation based on behavioral data as shown in Sect.~\ref{model} and \ref{experiment}.
This constitutes an approach for solving the metrological problem outlined in the Introduction. 

Objectivity, i.e. non-contextuality, is therefore not a primordial quality of numbers but merely a feature of the classical real-valued probability calculus limited to the description of subjective uncertainties about an independently existing reality. It is built in a set-theoretic Boolean algebra of events and a single-context Kolmogorovian probability space measured by real numbers \cite{Khrennikov2009b}. 

Subjective, contextual, semantic account of data, enabled by the complex-valued structure of cognitive Hilbert space as shown above, has no parallel in cognitive and behavioral models based on classical algebra of events and classical probability calculus, cf. \cite{UsoDomenech2019}.
Quantum theory thereby suggests a basis for the development of quantitative psychology and sociology bridging the divide between objective and subjective, quality and quantity, natural and human-centered sciences \cite{Brower1949,Howe1988,DeGrandpre2000,Cosmelli2008,Gelo2008,Gough2012,Lepskiy2018,Morf2018a,Kostromina2019}.


\paragraph{Semantic freedom and the feedback loop}

When an electron enters a Stern-Gerlach apparatus (Sect.~\ref{alternatives}), the quantum state of the system is prescribed by laws of quantum physics. Similarly to cognitive states ascribed to living systems, this state constitutes a semantic representation controlling the resolution of the corresponding objective uncertainty in a particular experimental context. 
Contrary to elementary systems studied in physics, semantics regulating behavior of higher animals is open to deliberate choice; just like signs acquire their meaning in the mind of a subject, in human cognition there is no such thing as the meaning of contexts per se \cite{Brier1998,Salvatore2020}. 

Subjectivity of semantics is reflected in the operation of the context representation algorithm described in Sect.~\ref{algorithm}: for a given set of behavioral probabilities, the returned representation is not unique but requires a prior setting of the normalization constant $N$ which crucially affects the result. 
Quantum cognitive triad thus can be considered as a tunable template used to organize individual behavior and subjective experience \cite{Krakauer2010}\footnote{In the spirit of QBism \cite{Fuchs2014,Khrennikov2016c,DeRonde2019}.}.
On a large scale, this results in a structure of enormous complexity - a unique personal worldview \cite{OConnor2009,Gabora2009a} involving many degrees of semantic freedom. Their subjective choice can be statistically assessed based on the efficiency of the resulting decisions which serves as a feedback signal for the learning system, cf. \cite{Litvintseva2009,Yukalov2009,Thompson2017b,Giannakis2019}.

\subsection{Quantum cognitive advantage}
\label{advant}

When a single behavioral context is considered in isolation, decision probabilities are defined solely by polar angle $\theta$ of the qubit cognitive state \eqref{eq5} so that azimuthal dimension $\phi$ is redundant.
Complex dimension of the probability amplitudes becomes necessary when behavior in multiple contexts has to be modeled without increasing dimensionality of the representation.
With representational space is restricted to two dimensions of the qubit Hilbert space, the decision probabilities observed in three different contexts require involvement of the complex-valued structure accounted above by azimuthal phase dimension.
In the above model, the corresponding parameters $\phi_{b}$ and $x$ define how representations of the contexts $a$ and $b$ are combined to form representation of the third context $c$. 

A similar situation realizes in physics where the complex-valued structure of Hilbert space is necessary to account for the outcome probabilities measured for the spin-1/2 systems in at least three binary observables \cite{Accardi1982,Khrennikov2003d}. Resulting accommodation of multiple contexts in a few-dimensional Hilbert space eliminating the need to compute multi-dimensional joint probability distributions is considered as a source of quantum computational advantage \cite{Khrennikov2019a}, cf.\cite{Thompson2017b,Giannakis2019}. 

This paper shows how the same computational principle applies to the cognition of living systems where many of behavioral contexts are represented in a single qubit cognitive space without increasing its dimensionality. Instead of accommodating every new decision context in a cognitive space of an enormous dimension, it is represented as a composition of related decision contexts that are already learned. This is a natural strategy for decision making under realistic behavioral restrictions \cite{Busemeyer2011}.

\subsection{Triad: a carrier of meaning}
\label{carrier}

Inspection of a single context (situation, alternative) is of no use; it is always the difference between the two that matters in practice. Besides, the inspection itself implies a reference frame which determines in what respect the two contexts are compared \cite{Popper1978}; this subjective <<point of view>> completes the minimal bundle of three entities enabling the meaning-based cognition of humans\footnote{This triple is not to be confused with three truth values addressed by ternary logic \cite{Toffano2020}.} \cite{Fisch1966,Brier1998,Levy2010,Martean2014a,Salvatore2020}.
Triadic nature of this subjectively-semantic representation of contexts is captured by the quantum cognitive triad model described in this paper. 

\paragraph{Triad versus dyad}

Triples of experimental contexts are requisite for a number of quantum phenomena. As noted in Sect.~\ref{advant}, it is the third context which does not fit in the two-dimensional real vector space asking for the complex-valued structure of the qubit Hilbert space; further, three observables are necessary to construct the Bell and Leggett-Garg inequalities used to identify exclusively quantum behavior of elementary systems \cite{Hess2016} as well as quantum-like cognitive contextuality \cite{Basieva2019}.
These and other phenomena modeled by quantum theory essentially manifest triadic features of reality which can be understood in semantic terms.
An example is given by predictive behavioral modeling via stable semantic relations encoded by azimuthal phase invariants which are essentially triadic quantities (Sect.~\ref{semstability}).


The benefit of triadic representation of information in comparison with a dyadic one is known in sociology where the difference between dyads and triads generates classification of organization types \cite{Simmel1950}. 
In particular, the rigid bond between elements of a bipartite systems contrasts with triadic relations favoring discrimination, deliberation, and reflexivity crucial for cognition and semantics \cite{Rustin1971,Martean2014a}; accordingly, the triadic structure is preferred to dyadic one in modeling of social relations \cite{Brashears2015} which constitutes a primal function of the intellect \cite{Humphrey1976,Vygotsky1979}.
The triad of a subject, a partner, and a reference object constitutes a minimal meaningful communicative situation \cite{Cornejo2008}.

Another advantage of triadic cognition compared to dyadic one, also observed in social structures, is its robustness to damage \cite{Simmel1950}. If contexts are represented in a association pairs then the erasure of any of them is irretrievable and fatal to the whole dyad. Any component of the triad, in contrast, is recoverable as a superposition of the other two based on stable semantic relations recorded in the quantum phase invariants akin to \eqref{eq8}.

\paragraph{Higher order structures}

The number of possible relations in tetrads, pentads, and more complex systems quickly grows beyond the limits of attention \cite{Miller1956,Holland1975,Goyal2016}; on the other hand, relations involving more than three sides are presumably reducible to triadic ones \cite{Mertz1979,Martean2014a}.
Given that, until more sophisticated probability measures \cite{Sorkin1994} become necessary, practical efficiency is likely to favor cognition of humans in terms of triadic structure considered in this paper.





\bibliographystyle{spmpsci}      
\bibliography{Triad}   

\end{document}